\documentclass[floatfix,showpacs,showkeys,amsmath,preprintnumbers,nofootinbib,onecolumn] {revtex4}
\usepackage{graphicx}
\usepackage{colordvi}
\usepackage{amsmath}
\usepackage{amssymb}
\usepackage{latexsym}
    \def\be{\begin{equation}}
    \def\ee{\end{equation}}
    \def\ba{\begin{eqnarray}}
    \def\ea{\end{eqnarray}}
\begin{document}

\title{Phantom energy accretion and primordial black holes evolution in Brans-Dicke theory}

\author{B. Nayak$^{*}$ and L. P. Singh$^{\dag}$}
\affiliation{{ Department of Physics, Utkal University, Vanivihar,
Bhubaneswar 751004, India}\\
E-mail: {$^*$bibeka@iopb.res.in and $^\dag$lambodar\_uu@yahoo.co.in}}
\begin{abstract}
In this work, we study the evolution of primordial black holes within  the context of Brans-Dicke theory by considering the presence of a dark energy component with a super-negative equation of state called phantom energy as a background. Besides Hawking evaporation, here we consider two type of accretions - radiation accretion and phantom energy accretion. We found that radiation accretion increases the lifetime of primordial black holes whereas phantom accretion decreases the lifespan of primordial black holes. Investigating the competition between the radiation accretion and phantom accretion, we got that there is an instant during the matter-dominated era beyond which phantom accretion dominates radiation accretion. So the primordial black holes which are formed in the later part of radiation dominated era and in matter dominated era are evaporated at a quicker rate than the Hawking evaporation. But for presently evaporating primordial black holes, radiation accretion and Hawking evaporation terms are dominant over phantom accretion term and hence presently evaporating primordial black holes are not much affected by phantom accretion. 
\end{abstract}
\maketitle
\section{Introduction}
Einstein's formulation of General Theory of Relativity(GTR) \cite{ein} in 1916
takes gravitational constant($G$) to be a time-independent quantity.
It is a pure tensor theory of gravity. Following Einstein's lead,
many scalar tensor theories have been developed as the extensions of GTR.
In all these theories  G is a time dependent quantity.
Among them Brans-Dicke(BD) theory \cite{bdt} is the simplest one.
In BD theory the gravitational constant is
set by the inverse of a time-dependent scalar field which couples to gravity
with a coupling parameter $\omega$. GTR can be recovered from BD theory in
the limit $\omega \to \infty$ \cite{bam}. BD theory also admits simple
expanding solutions \cite{mj} for scalar field $\phi(t)$ and scale
factor $a(t)$ which are compatible with solar system observations \cite{prg}.
BD theory is also sucessful in explaining many cosmological phenomena such as
inflation \cite{ls}, early and late time behaviour of the Universe \cite{ss},
cosmic acceleration and structure formation \cite{bm}, cosmic acceleration,
coincidence problem \cite{nb,bn} and problems relating to black holes \cite{bn1,bn2,bn3}.

Primordial Black Holes (PBHs) could be formed in the early universe due to various mechanisms, such as inflation \cite{cgl,kmz}, initial inhomogeneities \cite{carr}, phase transition and critical phenomena in gravitational collapse\cite{khopol,jedam}, bubble collision \cite{kss} or
the decay of cosmic loops \cite{polzem}
The formation masses of PBHs could be small enough for them to have evaporated
completely by the present epoch due to Hawking evaporation \cite{hawk}.
Early evaporating PBHs
could account for baryogenesis \cite{bckl,mds} in the universe. On the other
hand, longer lived
PBHs could act as seeds for structure formation \cite{mor} and could also form a significant component of dark matter \cite{blais}.

The finding of SN Ia observations \cite{agr} that the Universe is currently undergoing accelerated expansion constitutes the most intriguing discovery in observational cosmology of recent years. As a possible theoretical explanation, it is considered that the vacuum energy with negative pressure having equation of state $p=\gamma \rho$ termed as dark energy, is responsible for this acceleration. Some works \cite{bde, jq} have raised the possibility that the equation of state parameter $\gamma$ may be less than $-1$, which is known as phantom energy in literature. A peculiar property of cosmological models with phantom energy is the possibility of a Big Rip \cite{rrc}: an infinite increase of the scale factor of the Universe in a finite time. In the Big Rip scenario, the cosmological phantom energy density tends to infinity and all bound objects in the Universe are finally torn apart up to the subnuclear scales. The Present data on distant supernovas \cite{ua} showed that the presence of phantom energy with $-1.2<\gamma<-1$ in the Universe is highly likely. It has been already acknowledged that, being such an exotic physical species, the phantom energy may change the accretion regime of black holes.

In this work, we study evolution of PBH within a general phantom energy scenario. Taking phantom energy accretion along with Hawking evaporation and accretion of radiation, we show the PBH evolution in different cosmic era and discuss about the presently evaporating PBH.  
\section{PBHs in Brans Dicke theory}
For a spatially flat($k=0$) FRW universe with scale factor \textit{$a$} , the Einstein equations and the equation of motion for the JBD field $\Phi$ take the form
\be \label{bd1}
\frac{\dot{a}^2}{a^2}+\frac{\dot{a}}{a}\frac{\dot{\Phi}}{\Phi}-\frac{\omega}{6}\frac{\dot{\Phi}^2}{\Phi^2}=\frac{8\pi\rho}{3\Phi}
\ee
\be \label{bd2}
2\frac{\ddot{a}}{a}+\frac{\dot{a}^2}{a^2}+2\frac{\dot{a}}{a}\frac{\dot{\Phi}}{\Phi}+\frac{\omega}{2}\frac{\dot{\Phi}^2}{\Phi^2}+\frac{\ddot{\Phi}}{\Phi}=-\frac{8\pi p}{\Phi}
\ee
\be \label{bd3}
\frac{\ddot{\Phi}}{8\pi}+3\frac{\dot{a}}{a}\frac{\dot{\Phi}}{8\pi}=\frac{\rho-3p}{2\omega+3}  .
\ee

The energy conservation equation is
\begin{eqnarray} \label{ec}
\dot{\rho} + 3 \Big(\frac{\dot{a}}{a}\Big)({1+\gamma})\rho=0
\end{eqnarray}
on assuming that the universe is filled with perfect fluid describrd by equation of state $p=\gamma \rho$. The parameter $\gamma$ is $\frac{1}{3}$ for radiation dominated era$(t<t_e)$ and is $0$ for matter dominated era$(t>t_e)$, where time $t_e$ marks the end of the radiation dominated era $\approx 10^{11}$ sec.\\
Now equation (\ref{ec}) gives
\begin{eqnarray*}
\rho \propto \left\{
\begin{array}{rr}
a^{-4} &  (t<t_e)\\
a^{-3} &  (t>t_e)
\end{array}
\right.
\end{eqnarray*}
Barrow and Carr \cite{barrowcarr} have obtained the following solutions for $a$ and $G$ for different eras, as
\ba
a(t) \propto \left\{
\begin{array}{rr}
t^{1/2} &  (t<t_e)\\
t^{(2-n)/3} &  (t>t_e)
\end{array}
\right.
\label{sola}
\ea
and
\ba
G(t)= \left\{
\begin{array}{rr}
G_0\Big(\frac{t_0}{t_e}\Big)^n & (t<t_e)\\
G_0\Big(\frac{t_0}{t}\Big)^n & (t>t_e)
\end{array}
\right.
\label{solg}
\ea
where $t_0 \sim$ is the present
time, $G_0 \sim$ is the present value of $G$,
and $~n$ is a parameter related to $\omega$, i.e., $n=\frac{2}{4+3\omega}$ .
Since solar system observations \cite{bit} require that $\omega$ be large ($\omega \geq 10^4$), $n$ is very small ($n \leq 0.00007$) .
                        
From our previous work \cite{bn1}, we know that if we consider Hawking evaporation and accretion of radiation symultaneously, then the rate at which primordial black hole mass changes is given by
\begin{eqnarray}
\dot{M}_{PBH}=-\frac{a_H}{256\pi^3} \frac{1}{G^2M^2}+16\pi G^2M^2f\rho_r
\end{eqnarray}
where $a_H$ is the Stefan-Boltzmann constanat and $f$ is the accretion efficiency.

In a Universe also filled with phantom energy, the accretion of such exotic component should also be taken in account. Babichev, Dokuchaev and Eroshenko \cite{bde} have worked out a differential equation for a black hole accreting phantom energy only and found that phantom energy accretion decreases the overall black hole mass. The rate of accretion of phantom energy is given by
\begin{eqnarray}
\dot{M}_{ph}=16 \pi G^2M^2[\rho_{ph}+p(\rho_{ph})]
\end{eqnarray} 
But for phantom energy, $p(\rho)=\gamma \rho$ with $\gamma <-1$. So above equation takes the form
\begin{eqnarray}
\dot{M}_{ph}=16 \pi G^2M^2(1+\gamma)\rho_{ph}
\end{eqnarray} 

Considering the radiation accretion and Hawking evaporation term along with new phantom energy accretion term, one can get the PBH evolution equation as
\begin{eqnarray}
\dot{M}_{PBH}=-\frac{a_H}{256\pi^3} \frac{1}{G^2M^2}+16\pi G^2M^2f\rho_r+16 \pi G^2M^2(1+\gamma)\rho_{ph}
\end{eqnarray}
This equation is not exactly solvable but we solve it by using numerical methods for different cosmic era. \\
In our calculation, we have used $\gamma=-1.1$.\\

\section{Evolution of PBH in radiation-dominated era}
\subsection{Accretion of radiation}
The equation for accretion of radiation is
\begin{eqnarray}
\dot{M}_{rad}=16 \pi G^2M^2 f \rho_{r}
\end{eqnarray} 
where $\rho_r$ is the radiation energy density which varies with scale factor as $\rho_r=\rho_r^0 \Big(\frac{a}{a_0}\Big)^{-4}$. Using $\rho_r^0=\Omega_r^0 \rho_c$, above equation can be written as
\begin{eqnarray}
\dot{M}_{rad}=16 \pi G^2M^2 f \rho_{c} \Omega_r^0 \Big(\frac{a}{a_0}\Big)^{-4}
\end{eqnarray}
Again using equations (\ref{sola}) and (\ref{solg}), we get
\begin{eqnarray}\label{rrad}
\dot{M}_{rad}=16 \pi G_0^2\Big(\frac{t_0}{t_e}\Big)^{2n} M^2 f \rho_{c} \Omega_r^0 \Big(\frac{t}{t_e}\Big)^{-2}\Big(\frac{t_e}{t_0}\Big)^{(-8+4n)/3}
\end{eqnarray} 
By inserting numerical values of different quantities i.e. $G_0=6.67 \times 10^{-8}$ dyne-cm$^2$/gm$^2$, $\rho_c=1.1 \times 10^{-29}$ gm/cm$^3$, $t_0=4.42 \times 10^{17}$ sec along with $\Omega_r^0 \approx 10^{-5}$, one can find  
\begin{eqnarray} \label{ra}
\dot{M}_{rad}=1.94 f G_0 \Big(\frac{t_0}{t_e}\Big)^n\frac{M^2}{t^2}
\end{eqnarray}
On integration above equation gives 
\begin{eqnarray}
M=\frac{M_i}{1-1.94f\Big(\frac{t_i}{t}-1\Big)}
\end{eqnarray}
where $M_i$ is the initial mass of PBH formed at time $t_i$.\\
For large time $t$, the above equation asymptotes to
\begin{eqnarray}
M=\frac{M_i}{1-1.94f}
\end{eqnarray}
Thus for accretion to be effective, 
\begin{eqnarray}
f<\frac{1}{1.94}\approx0.515
\end{eqnarray}

The radiation accretion of a particular PBH having initial mass $M_i=10^{10}$ gm is shown in figure-1, which indicates  mass of the PBH increases with accretion efficiency. 

\begin{figure}[h]
\centering
\includegraphics{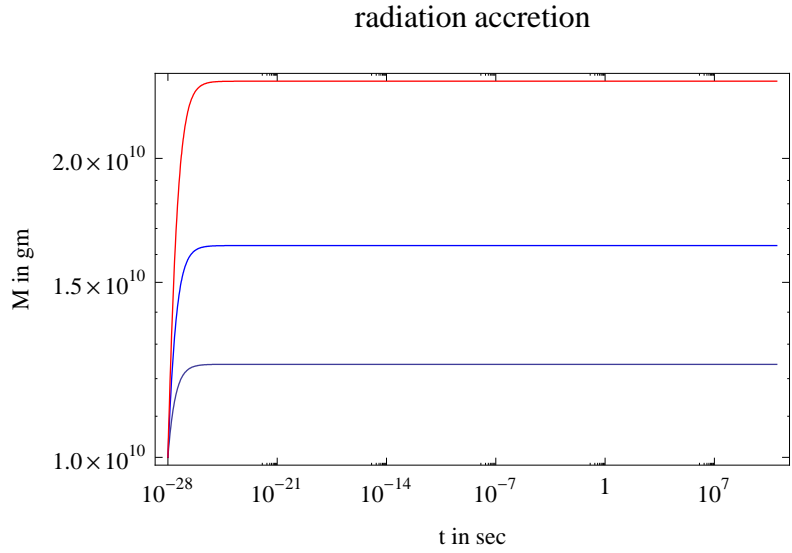}
\caption{Variation of PBH mass for diffent accretion efficiencies as $f=0.1,0.2,0.3$}
\label{fig1}
\end{figure}

For simpicity, all graphs are plotted in logarithmic scale.
\subsection{Accretion of phantom energy}
Accretion due to phantom energy is govern by the equation
\begin{eqnarray}
\dot{M}_{ph}=16 \pi G^2M^2(1+\gamma)\rho_{ph}
\end{eqnarray} 
But phantom energy density varies with scale factor as
\begin{eqnarray}
\rho_{ph}=\frac{\rho_{ph}^0}{|1+\gamma|}\Big(\frac{a}{a_0}\Big)^{-3(1+\gamma)}
\end{eqnarray} 
So phantom energy accretion equation becomes
\begin{eqnarray}
\dot{M}_{ph}=-16 \pi G^2M^2 \rho_{ph}^0 \Big(\frac{a}{a_0}\Big)^{-3(1+\gamma)}
\end{eqnarray} 
Again using equations (\ref{sola}) and (\ref{solg}), we get 
\begin{eqnarray} \label{rpha}
\dot{M}_{ph}=-16 \pi G_0^2 \Big(\frac{t_0}{t_e}\Big)^{2n} M^2 \Omega_{ph}^0 \rho_c \Big(\frac{t}{t_e}\Big)^{-3(1+\gamma)/2} \Big(\frac{t_e}{t_0}\Big)^{-(2-n)(1+\gamma)}
\end{eqnarray}

By solving equation (\ref{rpha}), Phantom energy accretion of a particular PBH having $M_i=10^{10}$ gm is shown in figure-2. This figure indicates phantom energy accretion is ineffective in radiation dominated era.
\begin{figure}[h]
\centering
\includegraphics{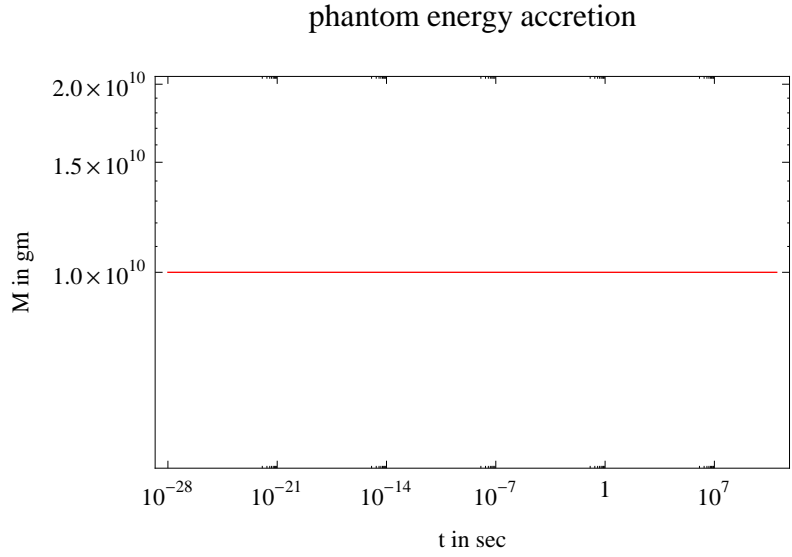}
\caption{Variation of PBH mass due to phantom energy accretion}
\label{fig2}
\end{figure}

\subsection{Complete evolution equation}

In radiation dominated era, the complete rate of change of PBH mass is given by
\begin{eqnarray} \label{rcee}
\dot{M}_{PBH}=-\frac{a_H}{256\pi^3} \frac{1}{G^2M^2}+16\pi G_0^2M^2f\rho_{c} \Omega_r^0 \Big(\frac{t}{t_e}\Big)^{-2}\Big(\frac{t_e}{t_0}\Big)^{((-8+4n)/3)-2n} \\ \nonumber -16 \pi G_0^2M^2 \rho_{c} \Omega_{ph}^0\Big(\frac{t}{t_e}\Big)^{-3(1+\gamma)/2}\Big(\frac{t_e}{t_0}\Big)^{-(2-n)(1+\gamma)-2n}
\end{eqnarray}

The complete evolution of a particular PBH having $M_i=10^{10}gm$ is shown in the figure-3. It is clear from the figure that evaporation time of a PBH delays due to accretion of radiation. More is the accretion, more delay is the evaporation.
\begin{figure}[h]
\centering
\includegraphics{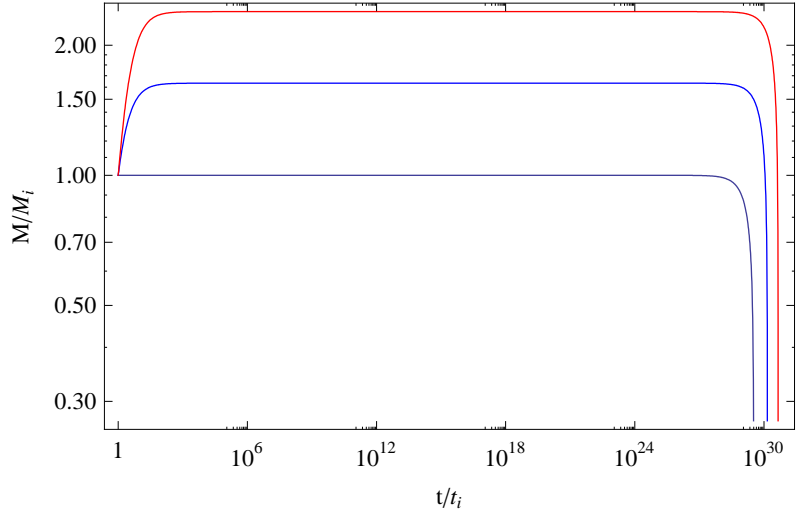}
\caption{Variation of evaporating time of a PBH in radition dominated era for $f=0, 0.2, 0.3$}
\label{fig3}
\end{figure}

\section{Evolution of PBH in matter-dominated era}
In this era, radiation accretion equation (\ref{rrad}) takes the form
\begin{eqnarray} \label{mr1}
\dot{M}_{rad}=16 \pi G_0^2M^2 f \rho_{c} \Omega_r^0 \Big(\frac{t}{t_0}\Big)^{((-8+4n)/3)-2n}
\end{eqnarray}
Integrating above equation, we get
\begin{eqnarray} \label{mr2}
M(t)=M_i\Big[1+\frac{48}{5+2n}\pi G_0 \rho_c \Omega_r^0 f \frac{t_0^{(8-n)/3}}{t_i^{(2-n)/3}} \Big\{\Big(\frac{t_i}{t} \Big)^{(5+2n)/3}-1 \Big\}\Big]^{-1}
\end{eqnarray}
For large time $t$, this equation asymptotes to
\begin{eqnarray} \label{mr3}
M(t)=M_i\Big[1-\frac{48}{5+2n}\pi G_0 \rho_c \Omega_r^0 f \frac{t_0^{(8-n)/3}}{t_i^{(2-n)/3}} \Big]^{-1}
\end{eqnarray}
\\
\\
In matter dominated era, phantom energy accretion equation (\ref{rpha}) takes the form
\begin{eqnarray} \label{matv1}
\dot{M}_{ph}=-16 \pi G^2 M^2 \Omega_{ph}^0 \rho_c \Big(\frac{t}{t_0}\Big)^{-(2-n)(1+\gamma)}
\end{eqnarray} 
Integating above equation, we get
\begin{eqnarray} \label{matv2}
M(t)=M_i\Big[1+\frac{16}{(2-n)|1+\gamma|+1-2n}\pi G_0 \rho_c \Omega_{ph}^0  \frac{t_i^{(2-n)|1+\gamma|+2-n}}{t_0^{(2-n)|1+\gamma|-n}} \Big\{\Big(\frac{t}{t_i} \Big)^{(2-n)|1+\gamma|+1-2n}-1 \Big\}\Big]^{-1}
\end{eqnarray}
For large time $t$, this equation gives
\begin{eqnarray} \label{matv3}
M(t)=M_i\Big[1+\frac{16}{(2-n)|1+\gamma|+1-2n}\pi G_0 \rho_c \Omega_{ph}^0  \frac{t_i^{(2-n)|1+\gamma|+2-n}}{t_0^{(2-n)|1+\gamma|-n}} \Big\{\Big(\frac{t}{t_i} \Big)^{(2-n)|1+\gamma|+1-2n}\Big\}\Big]^{-1}
\end{eqnarray}
The variation of PBH mass with time due to phantom energy is shown in figure-4 which indicates mass of the PBH decreases due to phantom energy accretion.
\begin{figure}[h]
\centering
\includegraphics{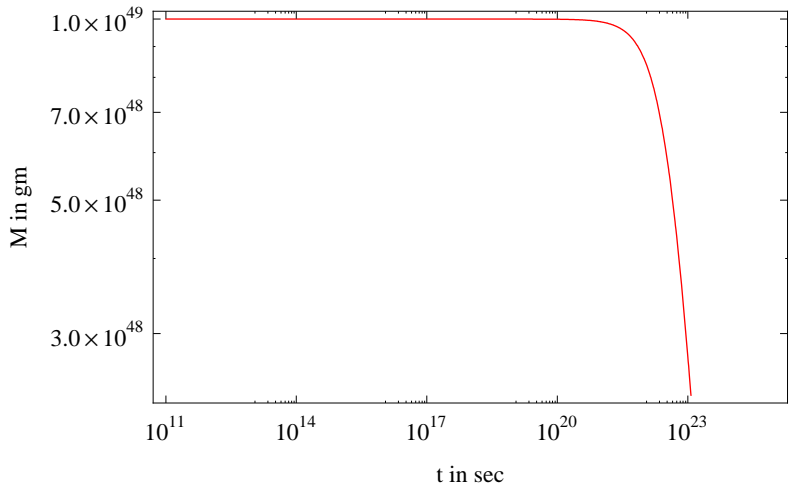}
\caption{Accretion of Phantom energy for $t_i=10^{11}$ sec is shown in the figure}
\label{fig4}
\end{figure}

The variation of PBH mass for phantom energy accretion and for radiation accretion is shown in figure-5. From this figure, it is clear that radiation accretion increases the PBH mass whereas phantom energy accretion decreases the PBH mass. 
 
\begin{figure}[h]
\centering
\includegraphics{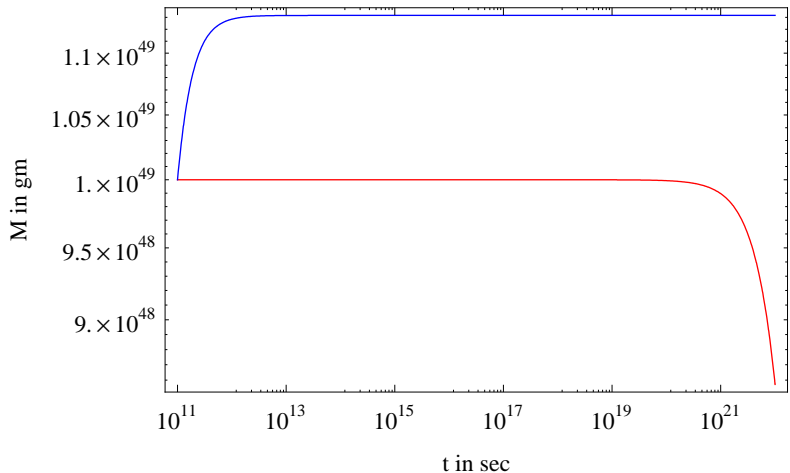}
\caption{Phantom accretion (Red) and Radiation Accretion (Blue) curves for $t_i=10^{11}$ sec are shown in the figure}
\label{fig5}
\end{figure}

Crossover from radiation accretion to phantom accretion occurs at a time $t_{eq}$ which can be calculated by equating the magnitude of two type of accreting masses. So from equations (\ref{mr3}) and (\ref{matv3}), one can get
\begin{eqnarray} \label{eq}
t_{eq}=t_i \times \Big[\frac{3\{(2-n)|1+\gamma|+1-2n\}}{5+2n} \Big]^{\frac{1}{(2-n)|1+\gamma|+1-2n}} \Big(\frac{f\Omega_r^0}{\Omega_{ph}^0}\Big)^{\frac{1}{(2-n)|1+\gamma|+1-2n}}
\end{eqnarray}
$t_{eq}$ is the time at which cross over from radiation to phantom occurs. This crossover time $t_{eq}$ increases with increase in accretion efficiency which is shown in figure-6. 
\begin{figure}[h]
\centering
\includegraphics{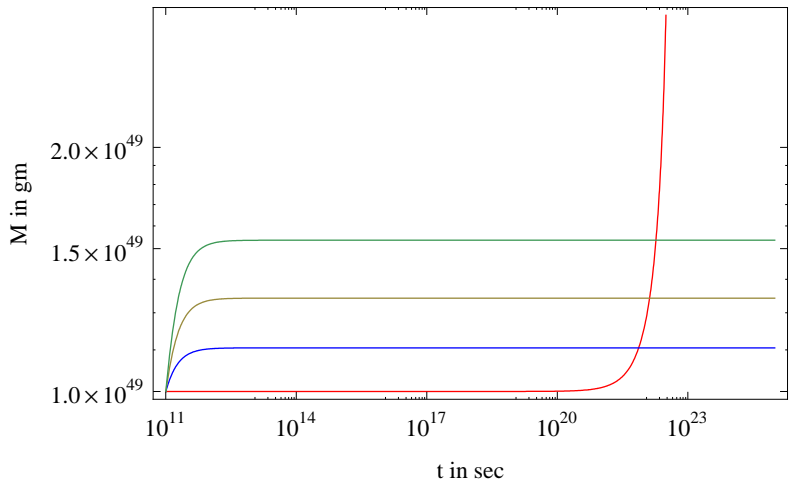}
\caption{Crossover from radiation to phantom accretion for $t_i=10^{11}$ sec is shown for different accretion efficiencies}
\label{fig6}
\end{figure}

In comparison with matter dominated era PBHs, rate of radiation accretion is much larger and rate of phantom accretion is much smaller for radiation dominated era PBHs. So for these PBHs the crossover time from radiation to  phantom ($t_{eq}$) comes much later and hence some radiation dominated era PBHs are completely evaporated without feeling the phantom energy accretion.     

In matter dominated era, the complete rate of change of PBH mass is given by
\begin{eqnarray} \label{mcee}
\dot{M}_{PBH}=-\frac{a_H}{256\pi^3} \frac{1}{G^2M^2}+16\pi G_0^2 \Big(\frac{t}{t_0}\Big)^{2n} M^2\rho_{c} \Big[f \Omega_r^0 \Big(\frac{t}{t_0}\Big)^{(-8+4n)/3}-\Omega_{ph}^0 \Big(\frac{t}{t_0}\Big)^{-(2-n)(1+\gamma)}\Big]
\end{eqnarray}
Solving equation (\ref{mcee}), we construct the Table-1. 
\begin{table}[h]
\begin{tabular}[c]{|c|c|c|}
\hline
\multicolumn{3}{|c|}{$t_{i}=10^{11}$ sec}\\
\hline
$f$  &  $t_{evap}$  &  $(t_{evap})_{ph}$  \\
\hline
$0$  &  $3.333\times10^{118}$sec  &  $4.445\times10^{43}$sec  \\
\hline
$0.2$  &  $7.379\times10^{118}$sec  &  $4.445\times10^{43}$sec  \\
\hline
$0.4$  &  $2.181\times10^{119}$sec  &  $4.445\times10^{43}$sec  \\
\hline
$0.6$  &  $1.211\times10^{120}$sec  &  $4.445\times10^{43}$sec  \\
\hline
$0.8$  &  $1.007\times10^{121}$sec  &  $4.445\times10^{43}$sec  \\
\hline
\end{tabular}
\caption{The evaporating times of the PBHs which are created at $t=10^{11}$ sec are displayed for several
accretion efficiencies for both cases without phantom accretion ($t_{evap}$) and with phantom accretion ($t_{evap})_{ph}$.}
\end{table}

From Table-1, we found that due to phantom energy accretion PBHs evaporated at a quicker rate than Hawking evaporation and their evaporating time is independent of radiation accretion.

\section{Evolution dynamics for presently evaporating PBH}
In this section, we discuss about the PBHs whose evaporating time is $t_0$. Solving equations (\ref{rcee}) and (\ref{mcee}) numerically, we construct the Table-2 for presently evaporating PBHs. 
\begin{table}[h]
\begin{tabular}[c]{|c|c|c|}
\hline
\multicolumn{3}{|c|}{$t_{evap}=t_0=4.42 \times10^{17}$sec}\\
\hline
$f$  &  $M_i$  &  $(M_i)_{ph}$  \\
\hline
$0$  &  $2.3669\times10^{15}$gm  &  $2.3669\times10^{15}$gm  \\
\hline
$0.1$  &  $1.908\times10^{15}$gm  &  $1.908\times10^{15}$gm  \\
\hline
$0.2$  &  $1.449\times10^{15}$gm  &  $1.449\times10^{15}$gm  \\
\hline
$0.3$  &  $0.989\times10^{15}$gm  &  $0.989\times10^{15}$gm  \\
\hline
$0.4$  &  $0.530\times10^{15}$gm  &  $0.530\times10^{15}$gm  \\
\hline
$0.5$  &  $0.714\times10^{14}$gm  &  $0.714\times10^{14}$gm  \\
\hline
\end{tabular}
\caption{The formation masses of the PBHs which are evaporating now are displayed for several
accretion efficiencies for both cases without phantom accretion ($M_i$) and with phantom accretion ($M_i)_{ph}$.}
\end{table}

It is clear from the Table-2 that phantom energy accretion does not affect lifetimes of presently evaporating PBHs because crossover time $t_{eq}$ from radiation to phantom is much larger than evaporating time. For same reason, the PBHs which are completely evaporated by present time are not influenced by the presence of phantom energy. 
\\
\\
Now we calculate the $\gamma$-ray constraint on PBHs by assuming observed $\gamma$-ray background arises due to presently evaporating PBHs.

The fraction of the Universes' mass going into PBHs at time $t$ is given by \cite{carr}
\begin{eqnarray}
\beta(t)=\Big[\frac{\Omega_{PBH}(t)}{\Omega_R}\Big](1+z)^{-1}
\end{eqnarray}
where $\Omega_{PBH}(t)$ is the density parameter associated with PBHs formed at time $t$, $z$ is the redshift associated with time $t$, $\Omega_R$ is the microwave background density.\\
If $M_*$ be the mass of the presently evaporating PBH, then initial mass fraction for that PBH becomes \cite{bn1}
\begin{eqnarray}
\beta(M_*) < \Big(\frac{M_*}{M_1}\Big)^{\frac{1}{2}} \times \Big(\frac{t_1}{t_0}\Big)^{\frac{(2-n)}{3}} \times 10^{-4}
\end{eqnarray}
The variation of $\beta(M_*)$ with $f$ drawn from variation of $M_*$ with $f$ is shown in the Table-3. The bound on $\beta(M_*)$ is strengthened as $f$ approaches its maximum value. But the bound is independent of phantom energy. Case is same for all other observed astrophysical constraints which arises due to completely evaporating PBHs.
\begin{table}[h]
\begin{tabular}[c]{|c|c|c|}
\hline
\multicolumn{3}{|c|}{$t_{evap}=t_0$}\\
\hline
$f$  &  $M_*$  &  $\beta(M_*) <$\\
\hline
$0$  &  $2.3669 \times 10^{15}$gm  &  $5.71 \times 10^{-26}$\\
\hline
$0.1$  &  $1.908 \times 10^{15}$gm  &  $5.13 \times 10^{-26}$\\
\hline
$0.2$  &  $1.449 \times 10^{15}$gm  &  $4.46 \times 10^{-26}$\\
\hline
$0.3$  &  $0.989 \times 10^{15}$gm  &  $3.69 \times 10^{-26}$\\
\hline
$0.4$  &  $0.530 \times 10^{15}$gm  &  $2.70 \times 10^{-26}$\\
\hline
$0.5$  &  $0.714 \times 10^{14}$gm  &  $0.99 \times 10^{-26}$\\
\hline
\end{tabular}
\caption{Upper bounds on the initial mass fraction of PBHs that are
evaporating today for various accretion efficiencies $f$.}
\end{table}
\section{Conclusion}
In this work, we study the evolution of primordial black holes within Brans-Dicke theory by considering the presence of a dark energy component with a super-negative equation of state called phantom energy as a background. Along with Hawking evapoartion, here we consider two type of accretions - radiation accretion and phantom energy accretion. We discuss each accretion term separately and then add all accretion and evaporation terms to study the complete evolution of PBH for different cosmic era. We found that in radiation dominted era phantom accretion is ineffective. But in matter dominated era, radition accretion increases the lifetime of PBHs whereas phantom energy accretion decreases the lifespan of PBHs. Investigating the competition between the radiation accretion and phantom accretion, we got that there is an instant during the matter-dominated era beyond which phantom accretion dominates over radiation accretion. So the PBHs which live beyond this transition time are affected by phantom energy accretion and evaporated at a quicker rate than their Hawking evaporation. Mainly those PBHs which are formed in the later part of the radiation dominated era and in matter dominated era are influenced by phantom energy. But for the presently evaporating PBHs, the radiation accretion and  Hawking evaporation terms are dominant over phantom accretion term. For these PBHs crossover  time from radiation to phantom comes much later than their evaporating time. Hence presently evaporating primordial black holes are not affected much by phantom accretion. Case is same for all primordial black holes which are evaporated till now. Thus, all observed astrophysical constraints on primordial black holes remain unaltered in the presence of phantom energy. 

\section*{Acknowledgements}
B.Nayak would like to thank the Council of Scientific and Industrial Research, Government of India, for the award of SRF, F.No. $09/173(0125)/2007-EMR-I$ .
                                                              

\end{document}